\documentclass[12pt,preprint,usenatbib,useAMS]{aastex}
\shorttitle{}
\shortauthors{P. Shalima et al.}

\begin{document}


\title{The Dust Content and Radiation Fields of Sample of Galaxies in the ELAIS-N1 Field}

\author{P. Shalima}
\affil{Indian Institute of Astrophysics, Bangalore-560034}
\author{Rupjyoti Gogoi, Amit Pathak} 
\affil{Department of Physics, Tezpur University, Napaam-784028}
\author{Ranjeev Misra, Ranjan Gupta}
\affil{Inter-University Center for Astronomy and Astrophysics, Pune-411007}
\and
\author{ D. B. Vaidya}
\affil{Gujarat College, Ahmedabad-380006}
\begin{abstract}
The Mid-IR colors ($F_{8}/F_{24}$) of galaxies together with their IR-UV luminosity correlations can be used to get some insight into the relative abundance of the different dust grain populations present in them. The ELAIS-N1 field contains thousands of 
galaxies which do not have optical spectra but have been observed in the 
Mid-IR by {\it Spitzer} and UV by {\it GALEX} making it ideal for these studies. As part of this work we have
selected a sample of galaxies from the ELAIS-N1 field which have photometric observations in the MIR and UV as well as  photometric redshifts from the SDSS database. We put the constraint that the redshifts are $\le$ 0.1, thereby giving us a total of 309 galaxies. We find that the majority of the galaxies in the sample are PAH dominated due to their high MIR flux ratio. We also find a reasonable correlation between the Mid-IR and the UV luminosities out of which the Mid-IR emission from PAHs at 8 $\mu$m is marginally better correlated than the 24 $\mu$m VSG emission with the UV luminosities. However, if we divide the sample based on their $F_{8}/F_{24}$ ratios which is also an indicator of metallicity, the MIR-UV correlation seems to increase with the $F_{8}/F_{24}$ ratio. But the MIR-UV correlations are not very different for the PAHs and the VSG population within the individual metallicity groups. 
\end{abstract}
\keywords{ISM, Galaxies}

\section{Introduction}
Most of the mid-infrared (MIR) emission (3-40 $\mu$m) from star-forming galaxies is due to thermal emission from dust grains heated by stellar
radiation.  The dust emission consists of two different components, i.e., broad emission features at 3.3, 6.2, 7.7, 8.6 and 11.3 $\mu$m
 from large aromatic molecules known as polycyclic aromatic hydrocarbons or PAHs  which are known to have vibrational transitions at these wavelengths
\citep{Leiger1984,All1985} and continuum emission from very small dust
grains (15-40$\sim$\AA, sometimes referred to as VSGs) \citep{Sellgren1984}.
The PAHs are excited by single photon heating and their Mid-IR luminosities are therefore independent of the radiation field density \citep{Draine2003}. The 8 $\mu$m emission from PAHs is correlated with cold dust emission at 160 $\mu$m \citep{Bendo2008}, making them good tracers of
the diffuse ISM in galaxies \citep{Peeters2004}.
Laboratory studies have shown that PAHs are excited mainly by UV photons (see \citet{Sellgren2001} and references therein) and there have been several observations of these emission features in regions of intense UV radiation.
However, the detection of these emission features in benign environments
with little UV radiation as well as in some early-type galaxies showed that
 they could also be excited by evolved stellar populations (for a recent review see \citet{Calzetti2011}). Such regions are dominated by large PAHs ($N_c>100$) where the absorption extends into the visible \citep{Li02}. The 8 $\mu$m emission from PAHs is
 found to be strong outside HII regions where individual PAH molecules can survive and they are irradiated by the general interstellar radiation field (ISRF)\citep{Gordon2008,Helou2004}.

The 24 $\mu$m continuum emission from VSGs (which may be PAH clusters or small carbonaceous grains), can only be excited
by highly luminous UV emitting stars and are found to be associated with the inner parts of HII regions.
\citet{Zhu2008} have found a tight correlation between the 24 $\mu$m (continuum emission from VSGs) and 70 $\mu$m luminosities both of which represent the hot dust.

The 8 $\mu$m emission from PAHs has been found to be tightly correlated with metallicity of the environment in which it is present \citep{Draine2007,Smith2007,Wu2007,Hunt2010}. This holds both globally and locally as in the case of HII regions \citep{Gordon2008,Wiebe2011,Khramtsova2013,Khramtsova2014}, where a tight correlation has been found between the $F_{8}/F_{24}$ ratios and the metallicities. While higher metallicity galaxies or regions lead to higher PAH abundances, lower metallicity objects, for eg., the Small Magellanic Cloud (SMC), tend to have much lower abundances with $\leq 0.4\%$ of the interstellar carbon residing in PAHs \citep{Li2002576}. This may be either due to lower rates of formation of PAHs in low metallicity objects or due to higher destruction. Low metallicity objects tend to have strong radiation field that easily destroys the PAHs except the very large ones or PAH clusters \citep{Madden2006,Rapacioli2006}. Thus, low metallicity environments with associated high radiation field are dominated by emission from VSGs compared to PAH emission. Several studies have found a reduction in the $F_{8}/F_{24}$ ratios in high radiation field regions which is lower than the predicted dust model values and 
have attributed it to the destruction of PAHs in intense radiation field regions or due to an increase in the VSG population due to the destruction of larger grains or both. However, the $F_{8}/F_{24}$ ratio is also found to be age dependent for different values of metallicities of HII complexes
\citep{Khramtsova2014}. Therefore the line and continuum emission observed in different intersteller environments are associated with different compositions of dust. 

Using a combination of the MIR emission from VSGs and PAH molecules together with the observed UV flux from stars in the galaxy, we intend to classify galaxies based on their $F_{8}/F_{24}$ ratios and understand the nature of dust present in them, in the absence of spectroscopic data.
Finally we compare our results with those obtained for other well known galaxy samples and HII regions.

\section{SAMPLE SELECTION}
We have used observations of the SWIRE ELAIS-N1 field ( RA=244$^{\circ}$, Dec=+54{$^\circ$}) from the Infrared Array Camera (IRAC: 3.6, 4.5, 5.8
\& 8.0 $\mu$m) and the Multi-Band Imaging Photometer (MIPS: 24 $\mu$m)
on the {\it Spitzer} \citep{Werner2004} and the far ultraviolet (FUV: {1350-1750 \AA}) and near ultraviolet (NUV: {1750-2800 \AA}) imagers on the {\it Galaxy Evolution Explorer} ({\it GALEX}; \citet{Martin2005}). At low redshifts, the IRAC
5.8 $\mu$m and 8.0 $\mu$m channels detect PAH emission features, with the 8.0 $\mu$m channel detecting it up to a redshift of 0.2 due to its larger bandwidth
\citep{Huang2007}. The other two IRAC channels 3.6 $\mu$m and 4.5 $\mu$m detect mostly stellar emission while the MIPS 24 $\mu$m band captures the thermal continuum emission of VSGs. We have selected those extended sources (Source
extractor ext flag $>=$ 2, \citet{Bert96}) that have been observed in all the IRAC channels, in the MIPS 24 $\mu$m band and the GALEX FUV and NUV bands. There are a total of  1169 extended sources in the field that have significant emission at the wavelengths considered here. We consider only those sources which have photometric redshift $\leq 0.1$ as observed by SDSS. With this restriction we are left with only 309 extended sources.
We have used the Kron fluxes from the SWIRE 2005 catalog for the  data \citep{Lonsdale2003} and the GALEX fluxes from the GALEX band merged catalogue corresponding to the tiles ELAIS-
N1-(00,03,12,01,02). The UV fluxes have been corrected for Galactic extinction using the E(B-V) values from the Spitzer catalog which originate from \citet{Schlegel}.
We have used the co-efficient of \citet{Helou2004} to correct for the stellar contamination in the 8 $\mu$m band.

\section{$F_{8}/F_{24}$ ratios and Dust model comparison}

\begin{figure}
\centering
\includegraphics[height=8cm,width=8cm]{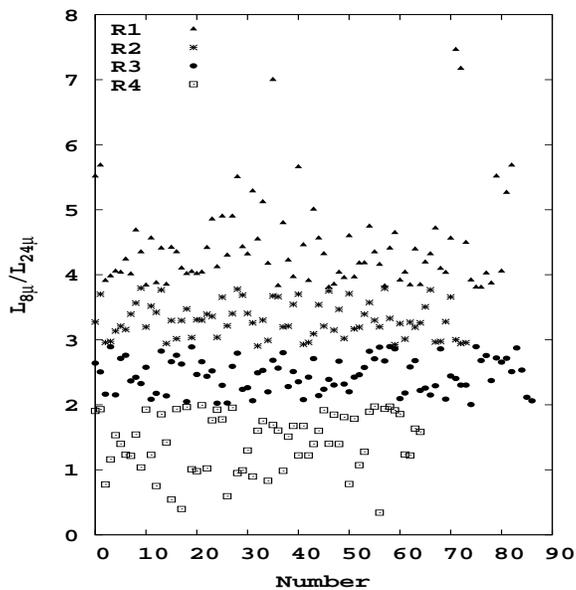}
\caption{Four groups of galaxies are represented by different symbols in the figure. The classification is done based on their $L_8/L_{24}$ ratio.}
\label{grp}
\end{figure}
We compared the sample of galaxies in the ELAIS-N1 field with SDSS observed sources and found 309 of them to have photometric redshifts. We then converted the redshifts to distances and computed Luminosities for these sources. We have plotted the $L_{8}/L_{24}$ ratio as a function of number of galaxies in Fig. \ref{grp}. We can see that majority of the galaxies have a high ratio indicating a higher PAH content in them. In order to study galaxies with different relative abundance of PAHs, we have divided the sample into four groups (R1-R4) based on their $L_{8}/L_{24}$ ratios. These groups can be considered to represent different metallicities since the $F_{8}/F_{24}$ ratios are found to be tightly correlated with galaxy metallicities. Galaxies belonging to different groups are marked by different symbols in Fig. \ref{grp}.

\begin{figure}
\centering
\includegraphics[height=8cm,width=8cm]{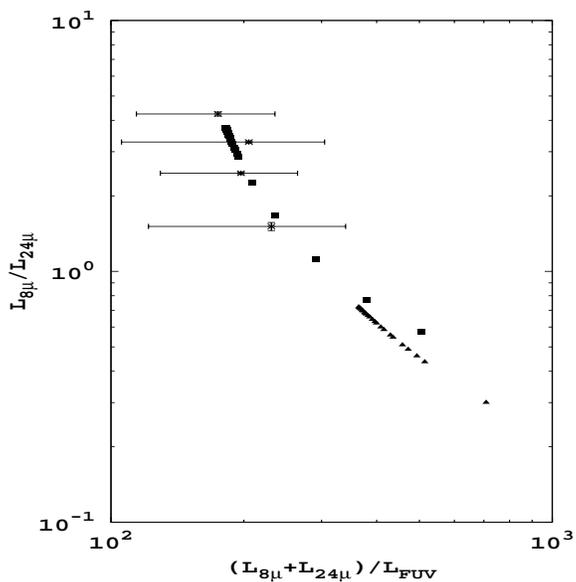}
\caption{\normalsize Logarithmic plot of the median of observed $\log(L_{8\mu}/L_{24\mu})$ vs. $\log(L_{8\mu}+L_{24\mu})/L_{FUV}$ (cross) and model flux ratios with U taken in place of FUV. The model values for qPAH = 4.58\% and qPAH = 0.47\% are plotted as boxes and triangles respectively.}
\label{8p24_fuv}
\end{figure}

\begin{figure}
\centering
\includegraphics[height=8cm,width=8cm]{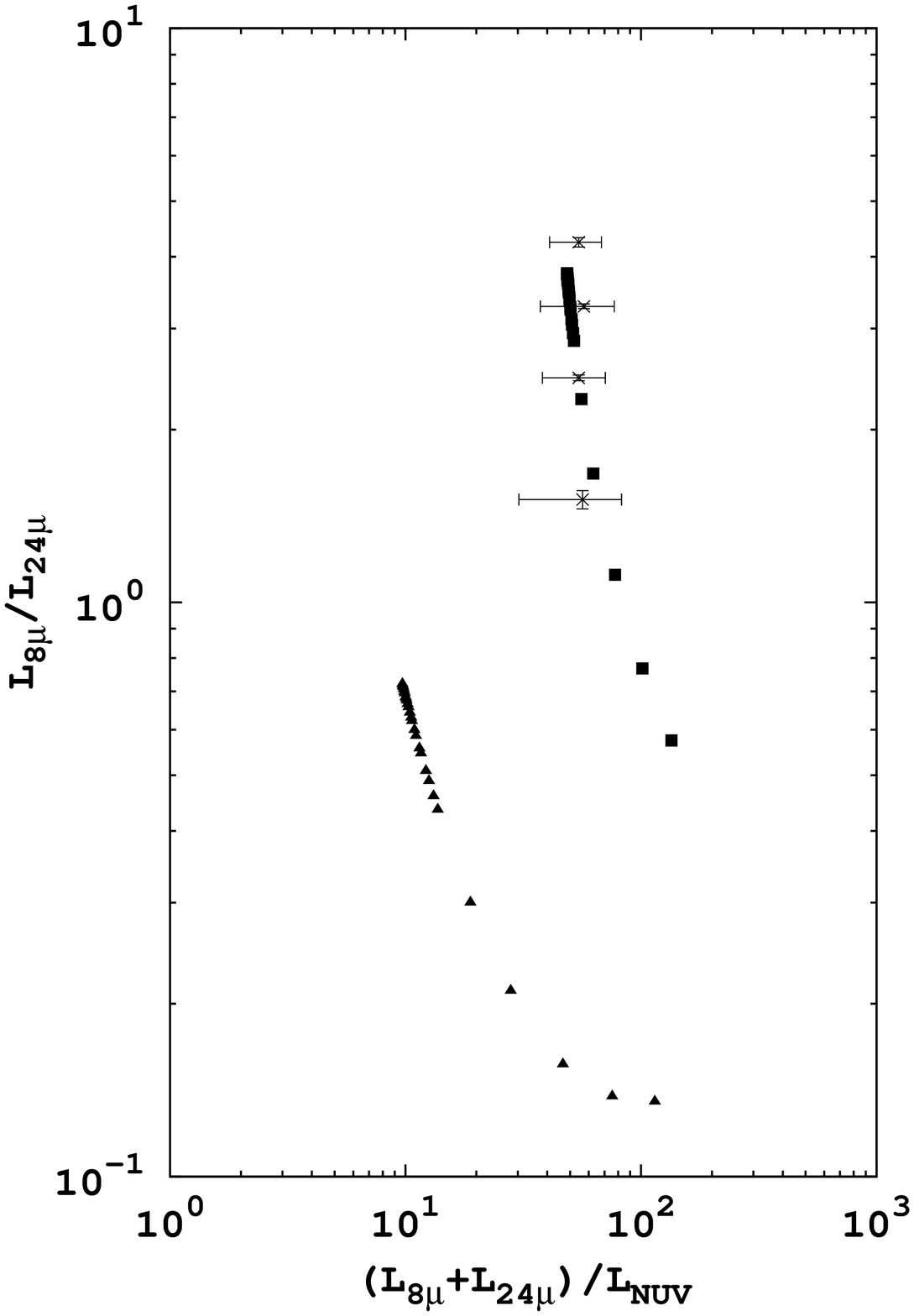}
\caption{\normalsize Logarithmic plot of the median of observed $\log(L_{8\mu}/L_{24\mu})$ vs. $\log(L_{8\mu}+L_{24\mu})/L_{NUV}$ flux ratios (cross) and model flux ratios with U taken in place of NUV. The model values for qPAH = 4.58\% and qPAH = 0.47\% are plotted as boxes and triangles respectively.}
\label{8p24_nuv}
\end{figure}
We have compared the observed $F_{8}/F_{24}$ ratios with the theoretical calculations of \citet{Draine2007}, where they have used a mixture of carbonaceous grains (comprised of PAHs and graphitic particles) and amorphous silicates exposed to radiation fields
ranging from $0.1-30000$ U, to predict mid-IR fluxes in each of the instrumental bands of Spitzer. U is the local interstellar radiation field in our Galaxy \citep{Mathis83}.
 For each radiation field model, the intensities are also computed for different
 values of the PAH mass present in grains denoted by qPAH, which is a function of the metallicity. 

We have plotted the median of the observed $L_{8}/L_{24}$ ratio vs. $L_{8+24}/L_{UV}$ ratio for each group (denoted by "x") together with the scaled model values, where U is taken in place of the UV fluxes in Figures \ref{8p24_fuv} and \ref{8p24_nuv}. The same values of U have been used for the FUV and NUV fluxes in order to compare the model with the observations. The model values for the two qPAH fractions 4.58\% and 0.47\% are marked as squares and triangles respectively.
Even though the local interstellar radiation field density, U has a spectrum which has a peak in the visible due to all the late type stars in the Solar neighbourhood, we can approximate it by the UV fluxes when studying the MIR flux ratios because the dust grains and PAHs are known to have very high extinction cross-sections in the UV compared to the optical. 
We see a reasonable fit between the model and observed profiles especially for the high PAH content. This may be an indication that the dust model with a high PAH fraction of 4.58\% is a fairly good representation of the dust content of these galaxies. Note that this value is representative of Milky Way dust as well.
\section{CORRELATION STUDIES}
As mentioned earlier, for 309 galaxies of our sample we have converted the photometric redshifts to distances and computed Luminosities. We studied the UV-IR luminosity correlations of our sample and looked for variations with metallicity. 
\begin{table}
\caption[]{\scriptsize Spearman-Rank Correlation co-efficients between FUV and MIR luminosities for the SDSS sample together with the probabilities. Correlations are also given for the different groups seperately.}
\label{Table6}
{\scriptsize
\begin{tabular}{cccccccccccc}
\hline
\hline
Band1&Band2&$R_{tot}$&Prob$_{tot}$&R$_{1}$&Prob$_{1}$&R$_{2}$&Prob$_{2}$&R$_{3}$&Prob$_{3}$&R$_{4}$&Prob$_{4}$\\
\hline
FUV&8.0&0.57&3.2$\times 10^{-28}$&0.60&1.5$\times 10^{-9}$&0.57&1.0$\times 10^{-7}$&0.55&2.7$\times 10^{-8}$&0.44&2.0$\times 10^{-4}$\\
FUV&24.0&0.55&5.4$\times 10^{-26}$&0.61&8.6$\times 10^{-10}$&0.57&1.2$\times 10^{-7}$&0.55&2.3$\times 10^{-8}$&0.42&4.3$\times 10^{-4}$\\
NUV&8.0&0.64&9.4$\times 10^{-37}$&0.66&8.5$\times 10^{-12}$&0.64&4.3$\times 10^{-10}$&0.62&1.0$\times 10^{-10}$&0.54&4.2$\times 10^{-6}$\\
NUV&24.0&0.62&1.8$\times 10^{-34}$&0.66&4.4$\times 10^{-12}$&0.64&5.2$\times 10^{-10}$&0.63&8.4$\times 10^{-11}$&0.51&1.1$\times 10^{-5}$\\
\hline\\
\end{tabular}
}
\label{tab3}
\end{table}

\begin{figure}
\centering
\includegraphics[width=6cm]{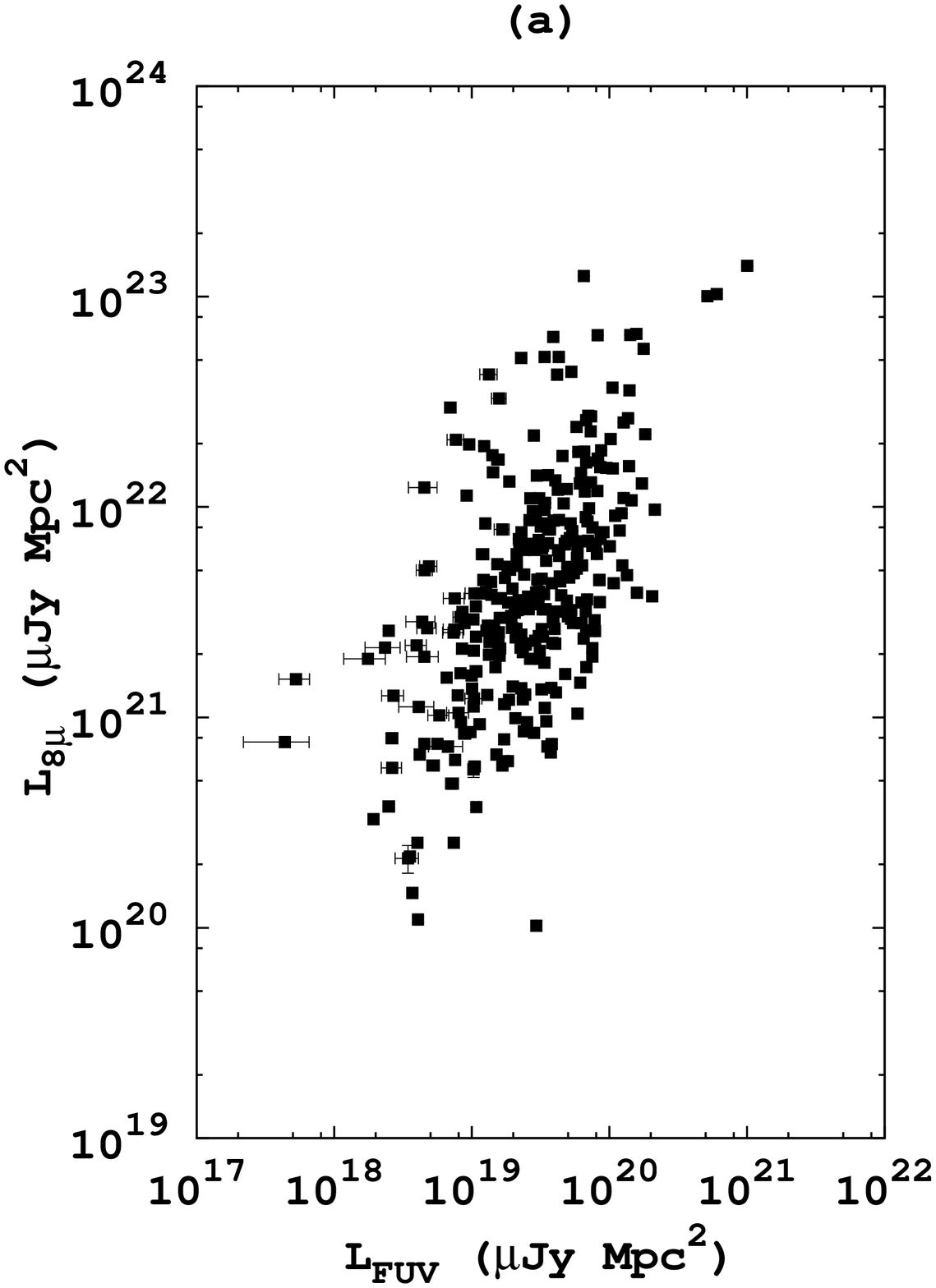}
\includegraphics[width=6cm]{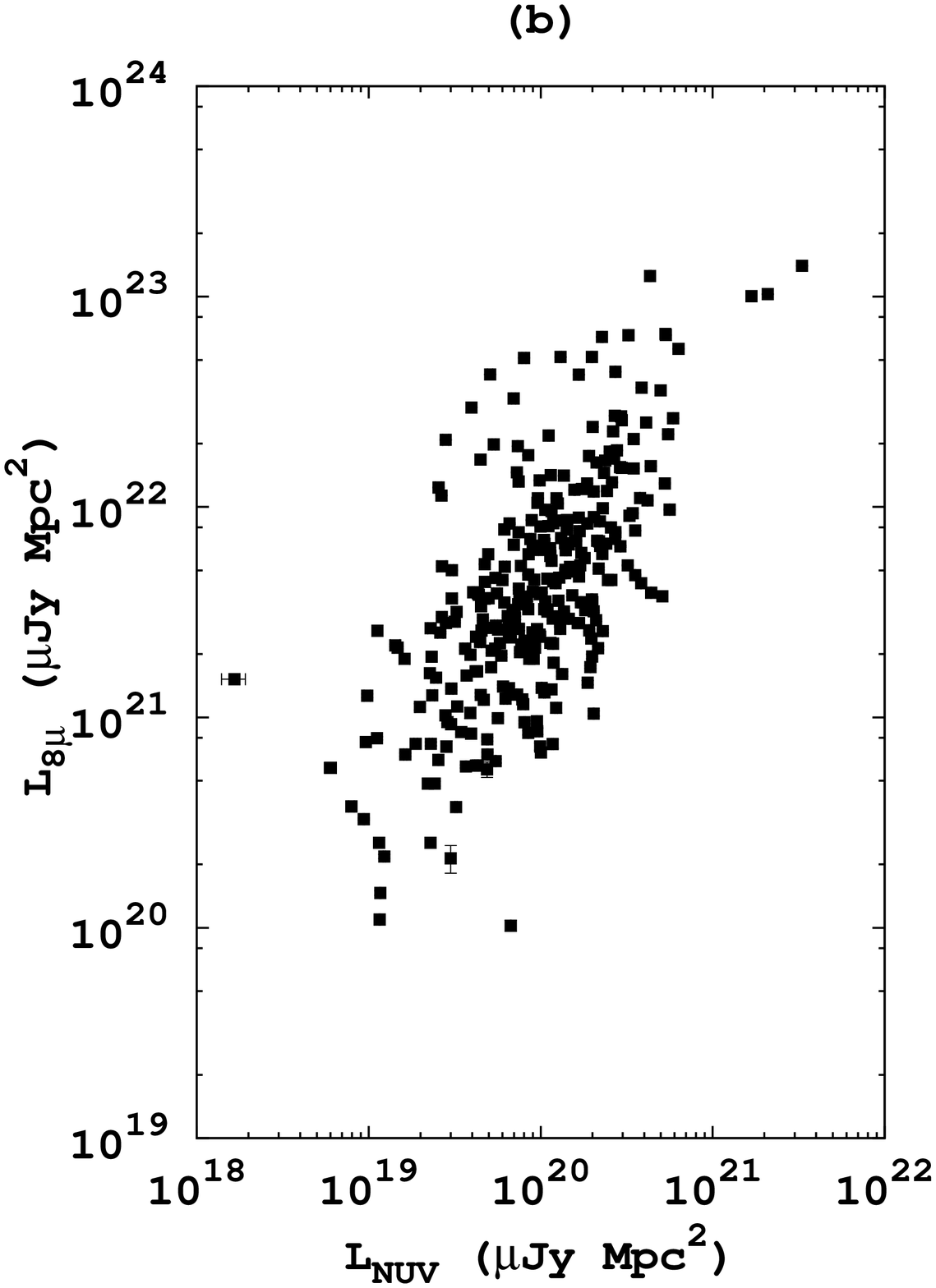}
\includegraphics[width=6cm]{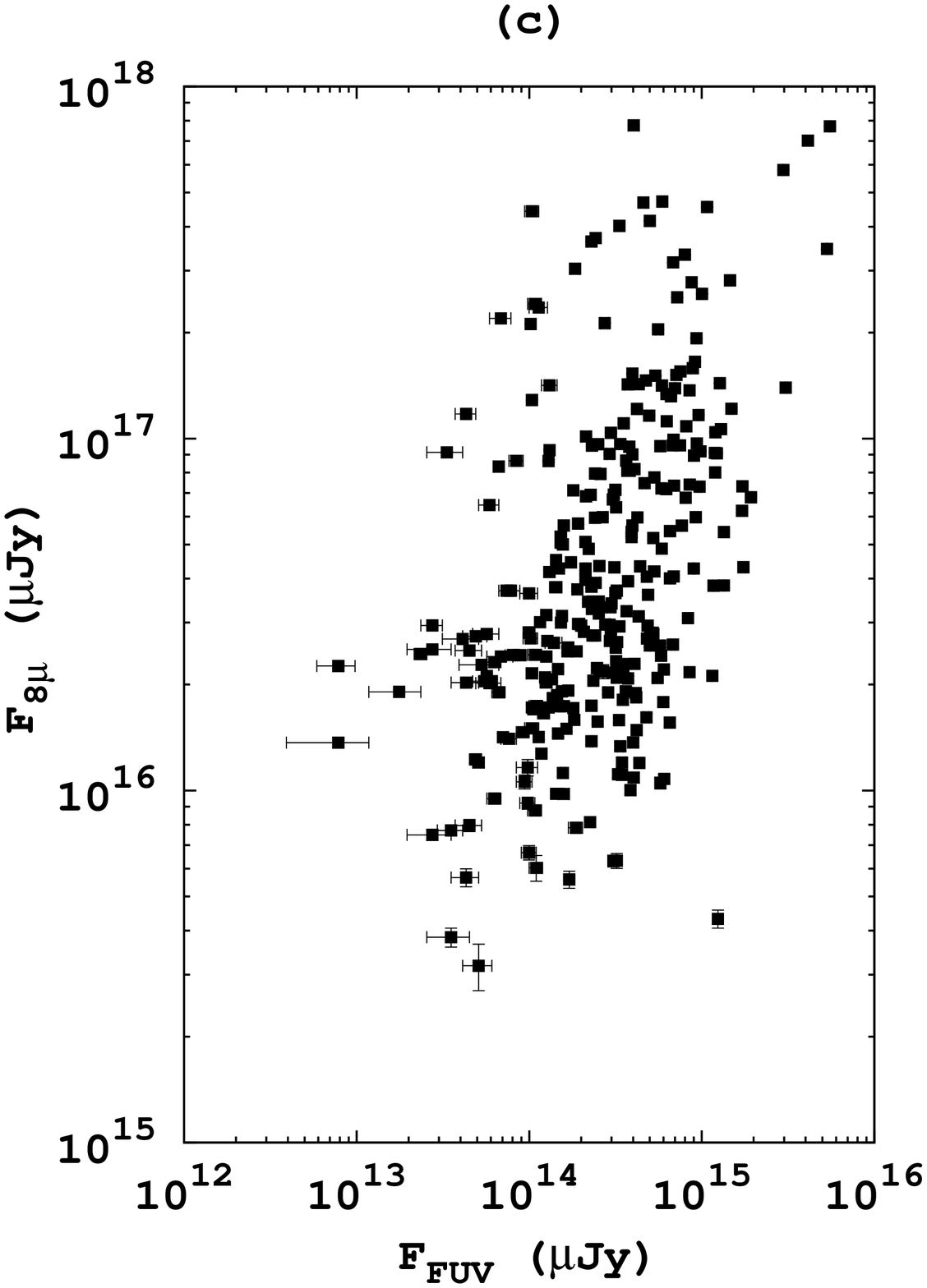}
\includegraphics[width=6cm]{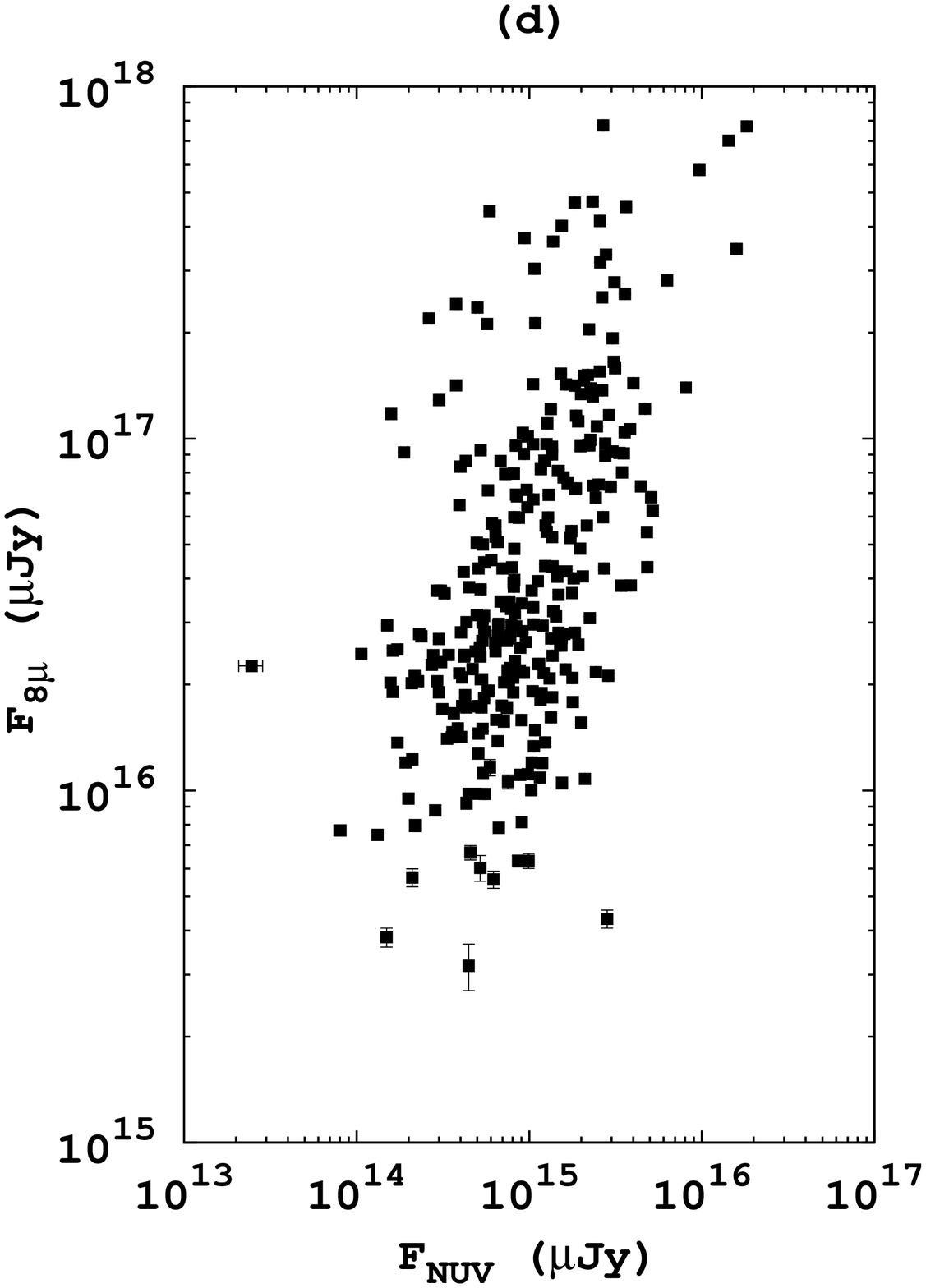}
\caption{\normalsize (a) and (b) represent the logarithmic plot of the observed UV vs 8 $\mu$m luminosities and (c) and (d) represent the logarithmic plot of the observed UV vs 8 $\mu$m fluxes.}
\label{L8fuv}
\end{figure}

\begin{figure}
\centering
\includegraphics[width=6cm]{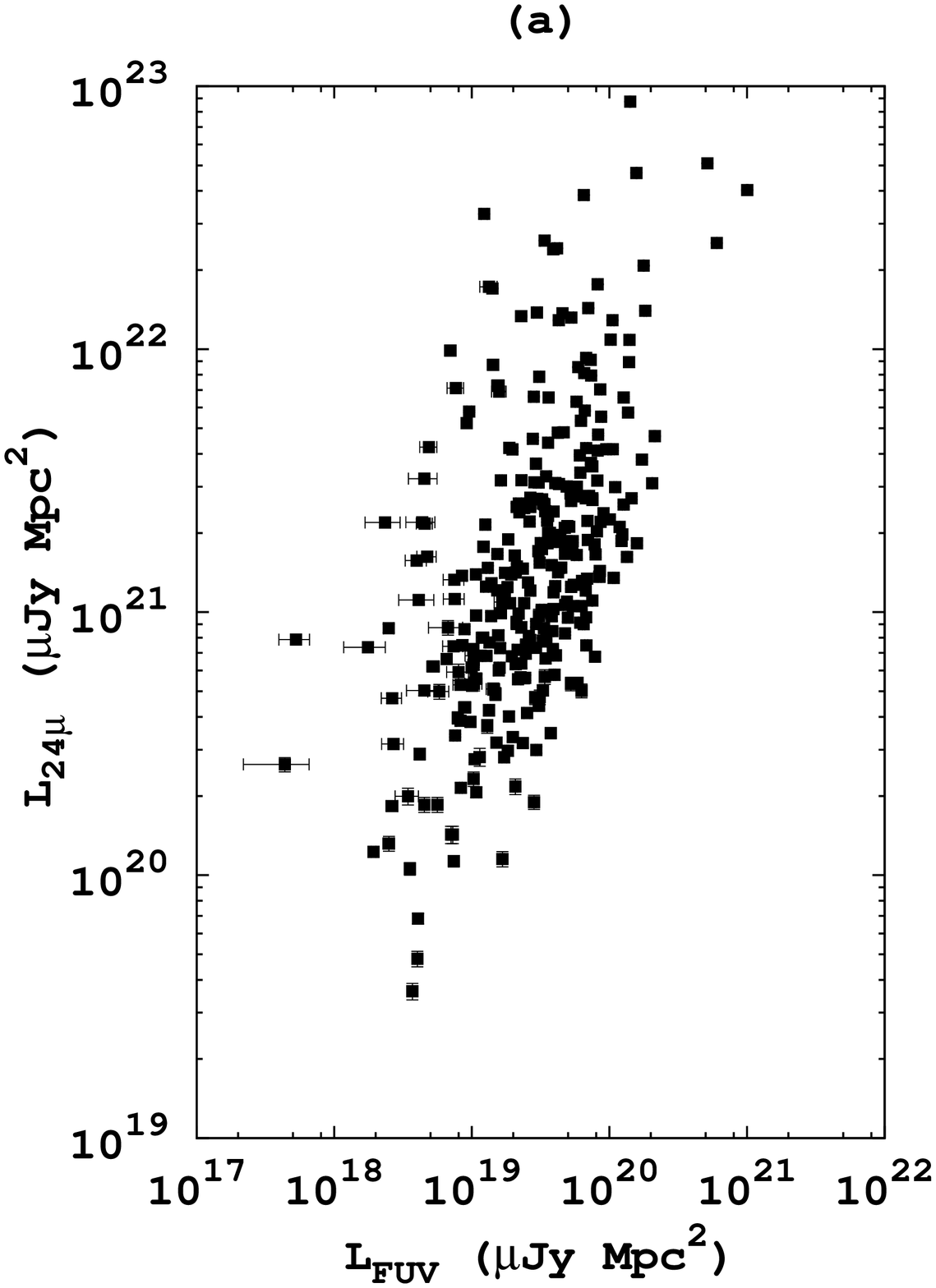}
\includegraphics[width=6cm]{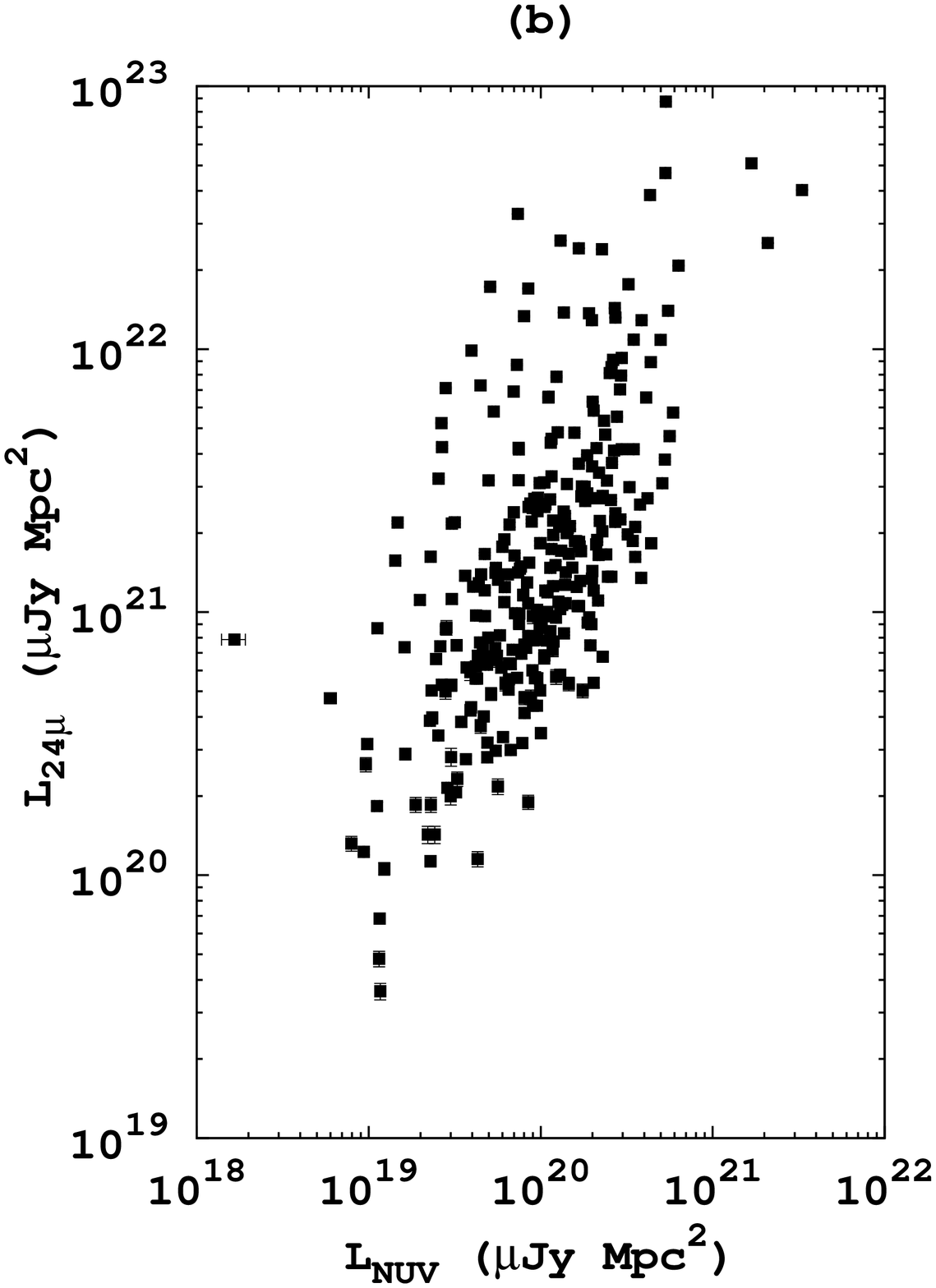}
\includegraphics[width=6cm]{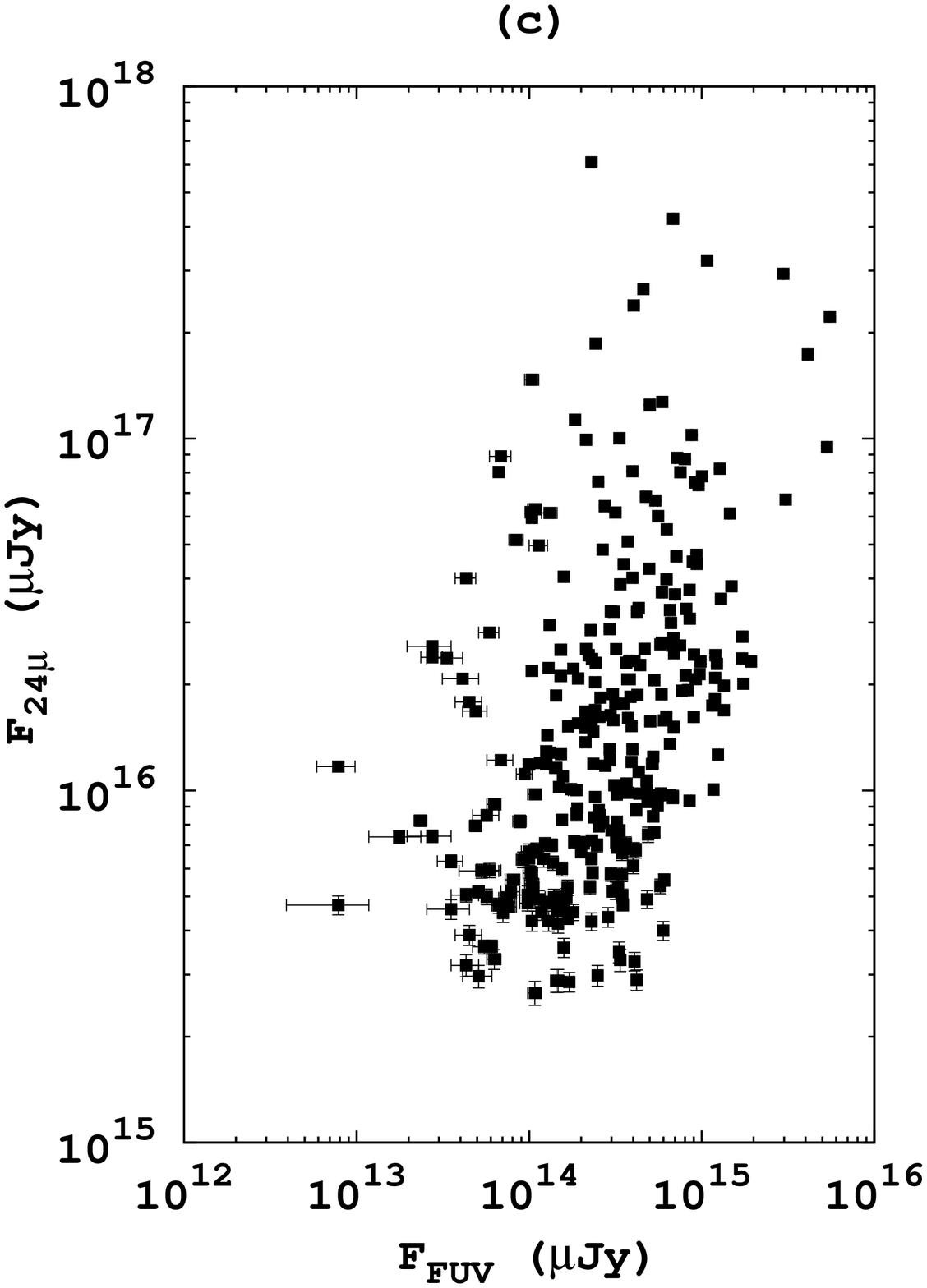}
\includegraphics[width=6cm]{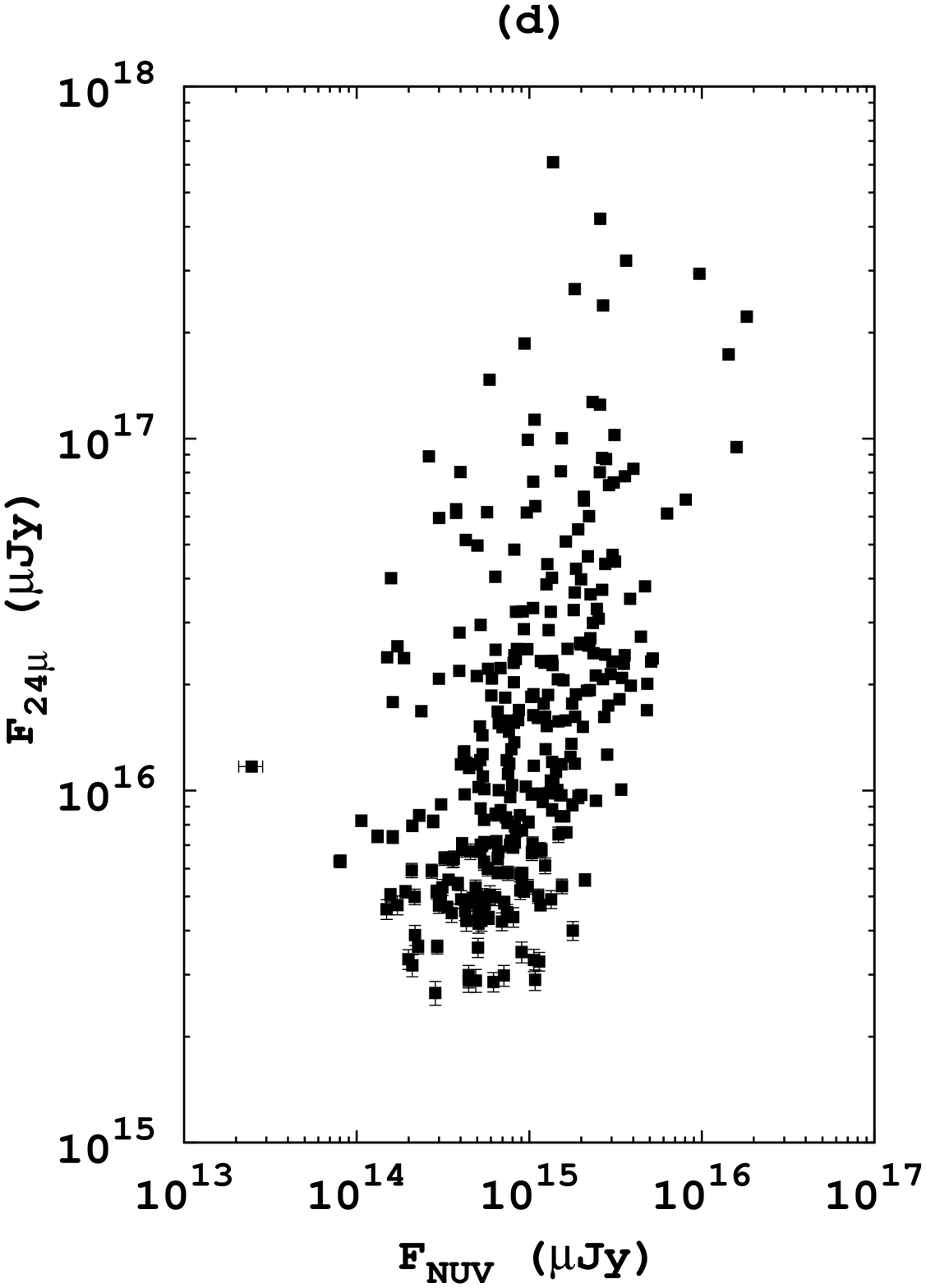}
\caption{\normalsize (a) and (b) represent the logarithmic plot of the observed UV vs 24 $\mu$m luminosities and (c) and (d) represent the logarithmic plot of the observed UV vs 24 $\mu$m fluxes.}
\label{L24nuv}
\end{figure}

We find  a good correlation between the UV and the 8.0 $\mu$m IRAC luminosity (see Fig. \ref{L8fuv} (a) and (b)), which is most sensitive to the PAH emission, with a better correlation for the NUV luminosities. Much of the scatter seen in the figures is probably due to differences in the intrinsic UV extinction, which will depend on the dust content, geometry and metallicity for each galaxy and hence difficult to correct for in this sample due to the unavailability of spectra. In Fig. \ref{L8fuv} ((c) and (d)) we have also shown the correlation between the UV and 8.0 $\mu$m IRAC flux.
 The 24 $\mu$m is also correlated with the UV luminosities as seen in Figure \ref{L24nuv}.
 The Spearman Rank Correlation co-efficients for these sources are tabulated in Table \ref{tab3}.  We see that the 8 $\mu$m emission is marginally better correlated with the FUV and NUV luminosities than 24 $\mu$m.
 This together with the higher $F_{8}/F_{24}$ ratio for the majority of the galaxies shows that the PAH emission dominates in these galaxies and the PAHs are mostly excited by the UV emitting stars present in them. Since the 24 $\mu$m feature requires VSGs to be heated by very hot massive stars, the slightly lower correlation together with the high $F_{8}/F_{24}$ ratios could indicate that in this sample, most of the galaxies do not have a large number of hot massive stars or have high metallicities as also seen from the comparison with the dust model.

 In order to verify if the correlation co-efficients vary with the metallicity,
we now compute the Spearman Rank correlations for the five $8\, \mu m/24\, \mu m$ groups seperately. We see that the correlation co-efficients reduce marginally with decrease in MIR flux ratios or metallicities i.e. as we go from group 1 to 4. 
But there is no difference in correlation between the PAH and UV emission or VSG and UV radiation.
This is in agreement with studies of starforming galaxies where the correlation co-efficients between MIR and UV luminosities are similar for PAH and VSG emission \citep{Zhu2008}.

\section{Conclusions}
We have used archival Spitzer and GALEX data of the integrated light of galaxies in the SWIRE ELAIS-N1 field in order to study and classify these galaxies on the basis of the type of dust grains in them.
We have used the  $F_{8}/F_{24}$ ratio to deduce the relative abundance of the PAH and VSG population. We find that majority of galaxies in the sample are PAH dominated.
Correlation studies of these galaxies show that that the 8 $\mu$m emission is reasonably correlated with the FUV and NUV radiation. This is also true for the 24 $\mu$m emission.
By considering the observed $F_{8}/F_{24}$ ratio as a proxy for the metallicity of the galaxy, we have divided the sample into four groups of different metallicities. We find that the observed MIR flux ratio follows the dust model ratio reasonably well especially for the higher PAH fraction.

Depending on the dominating dust component (PAH or VSG) that absorbs/scatters the UV radiation we could see a change in the MIR-UV correlations.
To check this, We have compared the MIR-UV correlations for the high and low metallicity galaxies, and find that the correlation co-efficients between the different groups reduce with reduction in the MIR flux ratio. For low metallicity galaxies, the harder radiation field \citep{Madden2006}, could be responsible for destruction of the PAH and VSGs thereby affecting the correlations. Or this could be just because of the higher extinction in these galaxies due to the larger dust content.
However, the correlation co-efficients are not very different for the PAH and VSG emission in each group, thereby making it difficult to conclude which grain population affects the UV radiation field in these galaxies. 
This could be because the abundance of the different grain populations is less sensitive to the UV radiation field than the metallicity. This has been seen in HII regions \citep{Khramtsova2014} and has been attributed to production of PAHs by aromatization of carbon grains in regions of high radiation field or low metallicities. Similarly the VSG population could also increase due to destruction of larger grains in these regions thereby influencing the MIR flux ratios.

Hence using this method we can see that even in the absence of spectroscopic measurements we can study the relative abundances of the different populations of grains in galaxies and the results are consistent with those obtained for well studied galaxies and HII regions with metallicity measurements from spectra.
However, it is not possible to identify what factor alters the relative abundance of the different grain populations in these galaxies,   
which probably means that several factors influence their relative abundance both globally and locally.
\acknowledgments
This work is based in part on observations made with the Spitzer Space Telescope, which is operated by the Jet Propulsion Laboratory, California Institute of Technology under a contract with NASA. This article is also based on data obtained by the NASA mission Galaxy Evolution Explorer(GALEX). PS would like to thank DST Fast Track programme(grant no.: SR/FTP/PS-68/2008) for financial support. RG would like to thank the IUCAA Associateship programme for their support and hospitality. AP thanks DST-SERB FAST TRACK programme and DST-JSPS project for financial support and IUCAA for visiting associateship.

\bibliography{spitzer}

\end{document}